\documentclass[reprint,amsmath,amssymb,aps]{revtex4-2}

\usepackage{xcolor}         
\usepackage{verbatim}
\usepackage{graphicx}
\usepackage{bm}
\usepackage{lineno}
\usepackage{setspace}

\begin{document}

\preprint{Physical Review Applied}

\title{Longitudinal-flexural wave mode conversion via periodically undulated waveguides with constant and graded profiles}

\author{Vin\'icius F. Dal Poggetto$^{1,2}$}
\email{vinicius.fonseca-dal-poggetto@univ-lille.fr}

\author{Fabio Nistri$^3$}

\author{Nicola M. Pugno$^{2,4}$}
\author{Marco Miniaci$^{1}$}
\author{Antonio S. Gliozzi$^3$}
\author{Federico Bosia$^3$}
\email{federico.bosia@polito.it}
 
\affiliation{
\footnotesize
 $^1$ Univ. Lille, CNRS, Centrale Lille, Junia, Univ. Polytechnique Hauts-de-France, UMR 8520 - IEMN - Institut d'Electronique de Micro\'electronique et de Nanotechnologie, F-59000 Lille, France\\
 $^2$ Laboratory for Bio-inspired, Bionic, Nano, Meta Materials \& Mechanics, Department of Civil, Environmental and Mechanical Engineering, University of Trento, 38123 Trento, Italy\\
 $^3$ Department of Applied Science and Technology, Politecnico di Torino, Corso Duca degli Abruzzi 24, 10129 Torino, Italy\\
 $^4$ School of Engineering and Materials Science, Queen Mary University of London, Mile End Road, London E1 4NS, United Kingdom
}

\begin{abstract}
Wave mode conversion allows to transform energy from one propagating wave type to another at a boundary where a change in material properties or geometry occurs.
Converting longitudinal waves to flexural ones is of particular interest in elasticity due to their significant displacement amplitudes, facilitating detection at the surface for practical applications.
Typically, the design of wave conversion devices requires
(i) the use of locally resonant structures with a spacing much shorter than the associated wavelengths, or (ii) architected media whose effective properties yield efficient mode conversion at selected frequencies. 
In both cases, the realization of these devices may incur in fabrication difficulties, thus requiring alternative solutions based on  simpler designs that can retain the wave manipulation capabilities of interest.
In this paper, we propose the use of single-phase periodically undulated beams to design phononic crystals that achieve wave mode conversion between longitudinal and flexural waves.
We derive the corresponding dispersion relations using the plane wave expansion method and demonstrate that the coupling between longitudinal and flexural wave modes can be manipulated using an undulated profile, generating mode veering with inverted group velocities.
The wave conversion mechanism is verified both computationally and experimentally, showing good agreement.
Our findings indicate a versatile design strategy for phononic crystals with efficient wave conversion property, enabling applications in structural health monitoring, sensing, and non-destructive testing.
\end{abstract}

\maketitle

\section{Introduction}

Wave mode conversion is a fundamental phenomenon in applications such as acoustic and ultrasonic sensing, yielding enhanced sensitivity by transforming one wave mode into another that is more easily detectable (e.g., longitudinal into shear waves)~\cite{auld1973acoustic,graff2012wave}.
This feature is particularly useful in non-destructive testing~\cite{wang2020non}, where mode conversion improves defect detection~\cite{rose2014ultrasonic,yeung2019time}, and also allows specific interaction with various materials, yielding structural defect discrimination through selective detection~\cite{prada2008local}.

Phononic crystals and elastic metamaterials have proven to be advantageous options for manipulating elastic and acoustic waves due to their tailorable properties and versatility~\cite{khelif2015phononic,laude2020phononic,craster2023mechanical,krushynska2023emerging}, with applications in sensing devices~\cite{jiao2023mechanical,pyo2024mechanical, miniaci2017proof},
energy harvesting~\cite{de2020graded,lee2023acoustic}, one-way transmission~\cite{gliozzi2019diode},
polarization control~\cite{de2021elastic},
topological waveguiding~\cite{laude2021principles}, and 
non-destructive testing~\cite{sukhovich2009experimental,sherwood20213d}.
While elastic metamaterials typically rely on locally resonant structures~\cite{liu2000locally}, phononic crystals are usually designed through the spatially periodic variation of material or geometrical properties.
As a consequence, phononic crystals present functionalities associated with Bragg scattering~\cite{page2004phononic,lu2009phononic}, leading to phenomena that occur at wavelengths comparable to the lattice spacing, thus limiting their operation in the subwavelength regime.
Beyond applications based on static quantities~\cite{zhang2017auxetic,mohammadi2021flexible}, mechanical metamaterials are mostly notable for their recent development in dynamic (i.e., frequency-dependent) applications.
Examples include acoustic metamaterials able to perform edge detection in the subwavelength range~\cite{moleron2015acoustic},
anisotropic graded-index metamaterials that achieve high signal-to-noise ratio for weak acoustic signal detection~\cite{chen2014enhanced}, and
cochlea-inspired metamaterials that enable the spatial discrimination of flexural waves based on their frequency content~\cite{dal2023cochlea}.


Elastic metamaterials have been widely employed in the design of wave mode conversion devices.
For instance, Colombi et al.~\cite{colombi2016seismic} proposed a system composed of a graded array of subwavelength resonators (''metawedge'') on an elastic substrate that can either (i) convert incident surface Rayleigh waves into bulk shear waves without reflections or (ii) reflect Rayleigh waves, creating a system with asymmetric transmission properties.
This concept was later studied by Chaplain et al.~\cite{chaplain2020tailored} to consider both longitudinal and flexural motion of the resonators and exploit Umklapp conversion.
Tian et al.~\cite{tian2022metamaterial} proposed a metamaterial consisting of a thin plate substrate decorated with resonating stubs to convert ultrasonic Lamb waves (A$_0$ and S$_0$ modes) into purely longitudinal shear horizontal waves (SH$_0$ mode), demonstrating complete conversion with a single type of unit cell while maintaining the propagation direction.
In another work, Wang et al.~\cite{wang2022total} utilized a ultra-thin metamaterial plate (thickness two orders smaller than the wavelength)
endowed with asymmetrical resonators whose anisotropy increases the coupling between the longitudinal/transverse motion to achieve nearly perfect mode conversion (above $95 \%$) between longitudinal and transverse waves.
Specifically in the context of longitudinal-flexural wave mode conversion, Cao et al.~\cite{cao2021elastic} proposed a system composed of a waveguide of periodic resonators whose Fano resonance hybridizes flexural and longitudinal waves, leading to highly localized bound modes~\cite{dal2024topological} with a theoretically infinite quality factor and perfect mode conversion. 
In all of these cases, the desired functionalities are achieved with the use of locally resonant structures that may be challenging to manufacture, due to size and geometrical constraints.

Metamaterials also provide excellent candidates to design wave conversion waveguides.
In~\cite{de2021selective}, De Ponti et al. designed a graded array of asymmetrically distributed pairs of resonators that yields mode conversion between incident flexural waves and reflected torsional waves. 
A similar concept was then applied to obtain selective mode conversion between torsional, flexural in-plane, and flexural out-of-plane waves, yielding group velocity inversion for a single wave type~\cite{iorio2024roton}.
Finally, Lemkalli et al.~\cite{lemkalli2024longitudinal} have demonstrated that longitudinal waves can be partially converted into torsional waves using a chiral unit cell whose edges are connected through tetragonal beams.
Therefore, introducing some sort of asymmetry in the design of the periodic unit cell seems to be a fundamental ingredient for the design of passive wave conversion waveguides.

Focusing on wave mode conversion using phononic crystals, a common strategy is to design unit cells whose effective medium properties present an anisotropic behavior, yielding coupling between distinct wave modes.
For instance, an anisotropic unit cell whose effective properties yield full conversion between longitudinal and shear waves in the ultrasonic regime was proposed in~\cite{yang2019monolayer}.
Full mode conversion was also achieved between longitudinal and shear modes in the ultrasound regime in~\cite{kweun2017transmodal}, where a transmodal Fabry-P\'{e}rot resonance mechanism is presented and investigated, deriving the conditions for broadband perfect mode conversion using an architected medium with tailored effective properties.
A similar concept was used to achieve longitudinal-torsional wave mode conversion employing a structured tubular waveguide in~\cite{yao2023elastic}.
However, designing phononic crystals that operate in the low-frequency regime and do not rely on locally resonant features remains an open challenge for wave mode conversion applications.
In particular, single-phase structures may enable designs that are amenable to fabrication, allowing versatile and low-cost wave conversion devices.

Phononic crystals in one- and two-dimensional structures with undulated profiles have been shown to be excellent options to achieve control over flexural waves.
Pelat et al. demonstrated this concept using corrugated profiles to manipulate the first Bragg band gap in beams~\cite{pelat2019control}, obtaining dispersion diagrams through the plane wave expansion (PWE) method.
This concept was later generalized to two-dimensional media using thin plate models (i.e., negligible shear strain and rotational inertia) with variable thickness to obtain low-frequency flexural band gaps with optimized widths by Dal Poggetto and Arruda~\cite{dal2021widening}.
Beyond purely flexural waves, folding-uncoupled origami metamaterial were also used to obtain band gaps based on coupled longitudinal-flexural behavior~\cite{xu2021coupled}. 
Likewise, Trainiti et al.~\cite{trainiti2015wave} investigated the occurrence of band gaps in undulated thin plates with asymmetric profiles (with respect to the plate midsurface) yielding a coupling between longitudinal and flexural motion.
In these cases, however, the correlation between the dispersion relations of undulated structures and the variation of wave modes associated with longitudinal or flexural behavior, which is fundamental for their exploitation as wave conversion devices, was not investigated.
Another particularity of interest concerning wave propagation in periodic media is the existence of sub-Bragg mechanisms (i.e., occurring below the first-order Bragg condition) observed in photonic crystals~\cite{huisman2012observation}.
Extending this concept to phononic crystals may allow subwavelength Bragg-like phenomena, with low-frequency wave manipulation capabilities~\cite{gzal2024sub,gzal2024subwavelength}.
In this work, we propose the design of beams with undulated profiles to perform mode conversion between longitudinal and flexural waves below the first-order Bragg condition.
This capability is achieved through a single-phase monolithic structure that can be easily manufactured, also enabling operation at various frequencies through the tuning of a single geometrical parameter.
Unlike the usual conversion mechanisms presented in the literature, our proposed structure also inverts the propagation direction, enabling wave conversion through reflection.
The corresponding dispersion curves are calculated considering non-negligible shear strain and rotational inertia, which improves their accuracy.
Our proposed wave conversion device is first justified analytically, then demonstrated numerically and validated experimentally.

This manuscript is divided as follows.
Section~\ref{sec_pwe_dispersion} presents a PWE formulation for the computation of dispersion diagrams in undulated beams considering normal-flexural coupling, shear strain, and rotational inertia.
In Section~\ref{sec_computation_dispersion} the proposed formulation is utilized to design a periodic unit cell with longitudinal-flexural coupling at a target operating frequency.
The numerical and experimental validation of the proposed wave conversion devices are then presented in Section~\ref{sec_wave_conversion}.
Conclusions and outlook for future work are outlined in Section~\ref{sec_conclusions}.

\section{Derivation of dispersion relations using the PWE method} \label{sec_pwe_dispersion}

\begin{figure*}[h]
 \centering
 \includegraphics{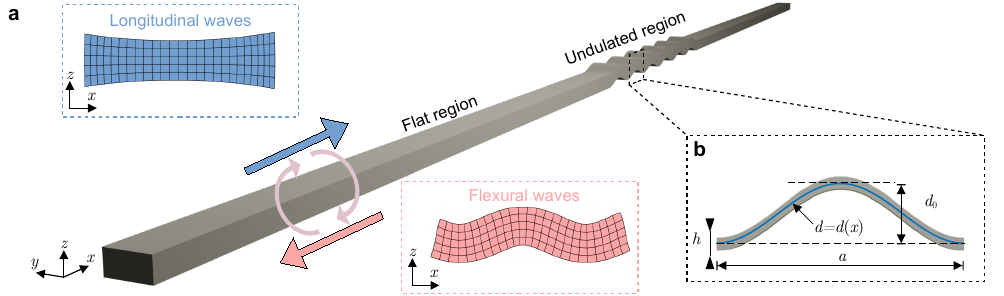}
 \caption{Longitudinal-flexural wave mode conversion device.
 \textbf{a}. The beam with an undulated profile is designed to convert incident longitudinal waves (blue arrow) to reflected flexural waves (red arrow).
 \textbf{b}. General profile $d(x)$ of the undulated region with lattice length $a$, constant thickness $h$, and undulation amplitude $d_0$.}
 \label{figure1}
\end{figure*}

Consider a beam whose centerline lies in the $x$-direction and presents an undulated profile in the $z$-direction, as illustrated in Fig.~\ref{figure1}\textbf{a}.
Although a simple sinusoidal function is considered for illustration purposes, the following derivation holds for any shape that can be expressed as a Fourier series.
The equations of motion relating the longitudinal ($u_x$) and transverse ($u_z$) displacements and the cross-section  rotation ($\psi_y$) with respect to its initial configuration in a beam with an undulated profile, considering both flexural and shear motion, 
are given by~\cite{chidamparam1993vibrations}
\begin{equation} \label{eq_dynamics}
 \begin{aligned}
  \frac{\partial N}{\partial x} + \kappa Q &= \rho A \frac{\partial^2 u_x}{\partial t^2} \, , \\
  \frac{\partial Q}{\partial x} - \kappa N &= \rho A \frac{\partial^2 u_z}{\partial t^2} \, , \\
  \frac{\partial M}{\partial x} + Q &= \rho I \frac{\partial^2 \psi_y}{\partial t^2} \, ,
 \end{aligned}
\end{equation}
where $x$ is the longitudinal beam coordinate, $t$ is the time coordinate, $N=N(x)$, $Q=Q(x)$, and $M=M(x)$ are, respectively, the internal longitudinal force, shear force, and flexural moment resultants, $\kappa=\kappa(x)$ is the beam curvature,
$\rho=\rho(x)$ is the specific mass density of the beam material, 
$A=A(x)$ and $I=I(x)$ are, respectively, the cross-sectional area and second moment of inertia of the beam with respect to its centerline.
The internal forces ($N(x)$ and $Q(x)$) and moment resultants ($M(x)$) are respectively given by~\cite{chidamparam1993vibrations}
\begin{equation} \label{eq_resultants}
 \begin{aligned}
  N(x) &= A E \bigg( \frac{\partial u_x}{\partial x} + \kappa u_z \bigg) \, , \\
  M(x) &= E I \frac{\partial \psi_y}{\partial x} \, , \\
  Q(x) &= \alpha A \mu \bigg( \frac{\partial u_z}{\partial x} - \kappa u_x - \psi_y \bigg) \, ,
 \end{aligned}
\end{equation}
where $E$ and $\mu$ are, respectively, the Young's modulus and shear modulus of the beam material, and $\alpha$ is a shear correction factor~\cite{timoshenko1972mechanics}. 
Combining Eqs. (\ref{eq_dynamics}) and (\ref{eq_resultants}) and assuming constant material properties and cross-section, we obtain
\begin{subequations}
 \begin{align}
  &AE \frac{\partial^2 u_x}{\partial x^2}
  - \alpha A \mu \kappa^2 u_x
  + A ( E + \alpha \mu ) \kappa \frac{\partial u_z}{\partial x} \nonumber \\
  + &AE \frac{\partial \kappa}{\partial x} u_z
  - \alpha A \mu \kappa \psi_y
  = \rho A \frac{\partial^2 u_x}{\partial t^2} \, , \\
  - &A ( E + \alpha \mu ) \kappa \frac{\partial u_x}{\partial x}
  - \alpha A \mu \frac{\partial \kappa}{\partial x} u_x
  + \alpha A \mu \frac{\partial^2 u_z}{\partial x^2} \nonumber \\
  - &A E \kappa^2 u_z
  - \alpha A \mu \frac{\partial \psi_y}{\partial x} 
  = \rho A \frac{\partial^2 u_z}{\partial t^2} \, , \\
  - &\alpha A \mu \kappa u_x
  + \alpha A \mu \frac{\partial u_z}{\partial x}
  + E I \frac{\partial^2 \psi_y}{\partial x^2} \nonumber \\
  - &\alpha A \mu \psi_y = \rho I \frac{\partial^2 \psi_y}{\partial t^2} \, .
 \end{align}
\end{subequations}

The computation of the dispersion relations $k=k(\omega)$ between the wavenumber $k$ and the propagating frequencies $\omega$ in a beam with a periodic curvature can be efficiently computed using the PWE method.
Following the methodology described in~\cite{dal2021flexural,dal2022wave}, we start by expressing the beam periodic curvature in the form of a Fourier series as
\begin{equation} \label{eq_kappa_series}
 \kappa(x) = \sum_{G_n} \, \widehat{\kappa}(G_n) e^{\text{i} G_n x} \, ,
\end{equation}
where $G_n = n \frac{2\pi}{a}$, $n \in \mathcal{Z}$, represents the reciprocal lattice vectors for a unit cell of length $a$, $\text{i} = \sqrt{-1}$ is the imaginary unit number, and
$\widehat{\kappa}(G_n)$ are its Fourier series coefficients, which can be computed through the inverse relation
\begin{equation}
 \widehat{\kappa}(G_n) = \frac{1}{a} \int_{0}^{a} \, \kappa(x) e^{-\text{i} G_n x} \, dx .
\end{equation}

Similarly, the periodic wavelength-dependent displacements and rotation ($u_x$, $u_z$, $\psi_y$) can be represented by time-harmonic Bloch waves~\cite{bloch1929quantenmechanik} expressed as
\begin{equation} \label{eq_disp_series}
 u(x,t) = e^{-\text{i} \omega t} \sum_{G_n} \, \hat{u}(G_n) e^{\text{i} (k+G_n) x} \, ,
\end{equation}
where $\hat{u} = \{ \hat{u}_x, \, \hat{u}_z, \, \hat{\psi}_y \}$ are the Fourier series coefficients corresponding to the spatial periodic function that composes the Bloch wave solution of $u_x$, $u_z$, or $\psi_y$. 

Equations (\ref{eq_kappa_series}) and (\ref{eq_disp_series}) can be substituted into Eqs. (\ref{eq_dynamics}) and (\ref{eq_resultants}), to yield the infinite system of linear equations given by
\begin{subequations} \label{eq_infinite_pwe}
\footnotesize
\begin{align} 
  & \sum_{G_j}
  \bigg[ E (k+G_j)^2 \delta_{i,j} + \alpha \mu \widehat{\kappa^2}(G_i-G_j) \bigg] \hat{u}_x(G_j) + \nonumber \\
  - & \sum_{G_j} 
  \bigg[ ( E + \alpha \mu ) \hat{\kappa}(G_i-G_j) \text{i} (k+G_j) 
  + E \widehat{\kappa'}(G_i-G_j) \bigg] \hat{u}_z(G_j) + \nonumber \\
  &+ \sum_{G_j} \bigg[ \alpha \mu \hat{\kappa}(G_i-G_j) \bigg] \hat{\psi}_y(G_j)
  = \omega^2 \rho \hat{u}_x(G_i) \, , \\
  & \sum_{G_j}
  \bigg[ ( E + \alpha \mu ) \hat{\kappa}(G_i-G_j) \text{i}(k+G_j) + \alpha \mu \widehat{k'}(G_i-G_j) \bigg]
  \hat{u}_x(G_j) +\nonumber \\
  + & \sum_{G_j} \bigg[ \alpha \mu (k+G_j)^2 \delta_{i,j} + E \widehat{\kappa^2}(G_i-G_j) \bigg]
  \hat{u}_z(G_j) +\nonumber \\
  + &\sum_{G_j} \bigg[ \alpha \mu \text{i}(k+G_j) \delta_{i,j} \bigg] \hat{\psi}_y (G_j)
  = \omega^2 \rho \hat{u}_z(G_i) \, , \\
  & \sum_{G_j} \bigg[ \alpha \mu \hat{\kappa}(G_i-G_j) \bigg] \hat{u}_x(G_j)
  - \sum_{G_j} \bigg[ \alpha \mu \text{i}(k+G_j) \delta_{i,j} \bigg] \hat{u}_z (G_j) +
  \nonumber \\
  + &\sum_{G_j} \bigg[ \frac{EI}{A} (k+G_j)^2 \delta_{i,j} + \alpha \mu \delta_{i,j} \bigg] \hat{\psi}_y(G_j)
  =\omega^2 \rho \frac{I}{A} \hat{\psi}_y(G_i) \, ,
\end{align}
\end{subequations}
where $\widehat{\kappa'}$ and $\widehat{\kappa^2}$ are, respectively, the Fourier series coefficients that represent the functions $\frac{\partial \kappa(x)}{\partial x}$ and $\kappa^2(x)$, $G_i$ ($G_j$) corresponds to the $i$-th ($j$-th) reciprocal lattice vector, and $\delta_{i,j}$ is the Kronecker delta (i.e, $\delta_{i,i}=1$ and $\delta_{i,j} = 0$ for $i \neq j$). For a given beam curvature $\kappa$, this system of equations can be used to obtain the propagating frequencies $\omega$ for each wavenumber $k$ of interest, thus yielding a dispersion relation.

Consider now a beam profile $d(x)$ corresponding to a single undulation described by
\begin{equation}
 d(x) = \frac{d_0}{2} \bigg( 1 - \cos \Big( \frac{2\pi}{a}x \Big) \bigg) \, , x \in [0,a] \, ,
\end{equation}
where $d_0$ is the profile height for a fixed beam thickness $h$, as shown in Fig.~\ref{figure1}\textbf{b}. The corresponding beam curvature is calculated as
\begin{equation}
 \kappa(x) = \frac{\partial^2 d(x)}{\partial x^2} = 
 d_0 \bigg(\frac{\pi}{a}\bigg)^2
 \bigg( e^{\text{i} \frac{2\pi}{a} x} + e^{-\text{i} \frac{2\pi}{a} x} \bigg) \, ,
\end{equation}
which can be immediately represented using a Fourier series in the form
\begin{equation} \label{eq_kappa_fourier1}
 \widehat{\kappa}(G_n) = d_0 \bigg(\frac{\pi}{a}\bigg)^2 ( \delta_{n,1} + \delta_{n,-1} ) \, .
\end{equation}

Analogously, the Fourier series coefficients for the terms $\frac{\partial \kappa(x)}{\partial x}$ and $\kappa^2(x)$ can be represented, respectively, as
\begin{subequations} \label{eq_kappa_fourier2}
\begin{align}
    \widehat{\kappa^\prime}(G_n) &= 2\text{i} \, d_0 \bigg(\frac{\pi}{a}\bigg)^3
    \bigg( \delta_{n,1} - \delta_{n,-1} \bigg) \, , \\
    \widehat{\kappa^2}(G_n) &= d_0^2 \bigg(\frac{\pi}{a}\bigg)^4
    \bigg( \delta_{n,2} + 2 \delta_{n,0} + \delta_{n,-2} \bigg) \, .
\end{align}
\end{subequations}

Substituting Eqs.~(\ref{eq_kappa_fourier1})--(\ref{eq_kappa_fourier2}) back into Eq.~(\ref{eq_infinite_pwe}) and truncating the number of plane waves by restricting the reciprocal lattice vector $G_n$ to $n \in [-N,N]$, $N \in \mathcal{N}$, corresponding to $1+2N$ plane waves, yields the eigenproblem
\begin{equation} \label{eq_pwe1}
 \left[
    \begin{array}{ccc}
     \mathbf{D}_{xx} & \mathbf{D}_{xz} & \mathbf{D}_{xy} \\
     \mathbf{D}_{zx} & \mathbf{D}_{zz} & \mathbf{D}_{zy} \\
     \mathbf{D}_{yx} & \mathbf{D}_{yz} & \mathbf{D}_{yy} \\
    \end{array}
 \right]
\left\{
    \begin{array}{c}
     \hat{\mathbf{u}}_x \\
     \hat{\mathbf{u}}_z \\
     \hat{\bm{\psi}_y} 
    \end{array}
 \right\}
 =
 \omega^2
 \left\{
    \begin{array}{c}
     \hat{\mathbf{u}}_x \\
     \hat{\mathbf{u}}_z \\
     \hat{\bm{\psi}_y} 
    \end{array}
 \right\} \, ,
\end{equation}
where the matrix terms are described in Appendix~\ref{app_eigenproblem_matrices} and the eigenvectors representing the Fourier series terms for the displacement and rotation components given by $\hat{\mathbf{u}} = \left\{ \begin{array}{ccccccc} \hat{u}(G_{-N}) & \cdots & \hat{u}(G_{-1}) & \hat{u}(G_{0}) & \hat{u}(G_{1}) \cdots & \hat{u}(G_N) \end{array} \right\}^T$, for $u = \{ u_x, u_z, \psi_y \}$.

\section{Computation of dispersion relations and experimental verification} \label{sec_computation_dispersion}

In this section, we apply the derived semi-analytical method to obtain the dispersion relations considering a homogeneous isotropic medium made of Aluminum, with Young's modulus $E = 70$ GPa, Poisson's ration $0.33$, and specific mass density $\rho = 2700$ kg/m$^3$.
The cross section of the beam is rectangular, with width $b=4$ mm and height $h=2$ mm. The unit cell length is fixed as $a=10$ mm, which was chosen due to fabrication constraints for the total length of the test specimen ($300$ mm).
Then, we utilize an experimental set-up to verify the existence of the computed band gaps.

\subsection{Computation of dispersion relations}

\begin{figure*}[h!]
 \centering
 \includegraphics{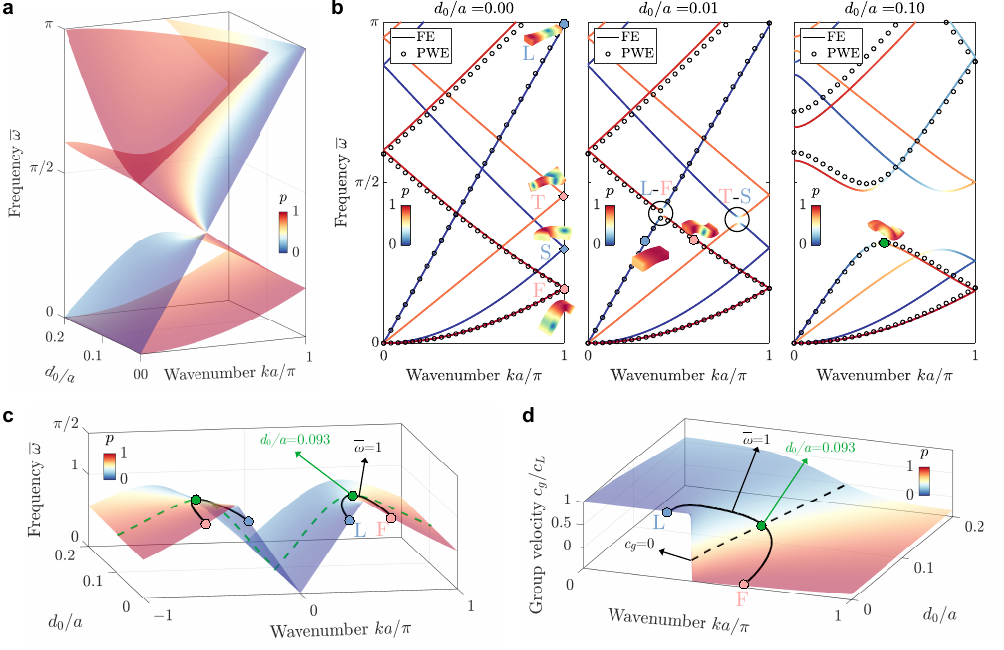}
 \caption{Dispersion relations computed using the PWE and FE as a function of the normalized undulation height $d_0/a \in [0,0.2]$ in the frequency range $\overline{\omega} \in [0,\pi]$.
 \textbf{a}. Dispersion curves computed with the PWE method with the out-of-plane polarization metric ($p$) ranging from $0$ (purely longitudinal, blue) to $1$ (purely transverse, red).
 \textbf{b}. Comparison of dispersion curves obtained using the PWE (black circles) and FE (colored lines) for increasing values of $d_0/a$.
 For $d_0/a=0$ (left panel), highlighted branches are labeled as L (longitudinal), F (flexural), T (torsional), and S (shear), with corresponding wave modes shown at the edge of the first Brillouin zone ($ka/\pi = 1$).
 For $d_0/a=0.01$ (middle panel), the longitudinal and flexural branches present veering (L--F) due to the coupling between in-plane ($p<0.5$) and out-of-plane ($p>0.5$) branches. A similar behavior is observed for the torsional and shear branches (T--S), only captured by the FE model.
 For $d_0/a=0.10$ (right panel), a significant band gap is noticed between the coupled L--F (T--S) branches, with a polarization $p \approx 0.5$ at the local maximum of the coupled branch.
 \textbf{c}. Isofrequency line for $\overline{\omega}=1$ (black line) indicated on the second dispersion branches of (\textbf{a}), with a maximum value (green circle) obtained at $d_0/a = 0.093$ (green dashed line). The blue and red circles correspond, respectively, to the longitudinal and flexural wave modes at a flat unit cell ($d_0/a=0$).
 \textbf{d}. Normalized group velocity ($c_g/c_L$) computed along the isofrequency line $\overline{\omega}=1$ (black line), indicating a zero group velocity ($c_g=0$) at $d_0/a = 0.093$.
 }
 \label{figure2}
\end{figure*}

We initially use the PWE formulation given by Eq.~(\ref{eq_pwe1}) to compute the dispersion relations corresponding to the unit cell with increasing values of undulation $d_0$. For these computations, we utilize $11$ plane waves, which present good convergence due to the simple harmonic expression of the chosen undulation profile.
The computed dispersion relations (frequency $\omega$ vs. wavenumber $k$) are presented in Fig.~\ref{figure2}\textbf{a} for increasing values of $d_0/a \in [0,0.2]$, where we also include in colour scale a polarization metric given by $p(\hat{\mathbf{u}}_x,\hat{\mathbf{u}}_z) = \sum_G |\hat{u}_z|^2 / (\sum_G |\hat{u}_x|^2 + |u_z|^2 )$, thus indicating $p=0$ for purely longitudinal ($u_z=0$) and $p=1$ for purely transverse ($u_x=0$) motion.
We also utilize the normalized frequency $\overline{\omega} = \omega a /c_L$, where $c_L = \sqrt{E/\rho}$ is the (quasi-)longitudinal wave speed in the beam material, restricting the results to $\overline{\omega} \in [0,\pi]$.
This frequency normalization is useful to represent the ratio between the unit cell length $a$ and the longitudinal wavelength $\lambda_L$, obtained by recalling that $k_L = \omega/c_L = 2\pi/\lambda_L$, where $k_L$ is the longitudinal wavenumber, which leads to the relation $\overline{\omega}/2\pi = a / \lambda_L$.

We then numerically validate the dispersion relations obtained using the PWE method  with the ones obtained using a finite element (FE)-based solution enforcing periodic boundary conditions~\cite{mace2008modelling}.
For the FE computations, we use hexahedral linear solid elements~\cite{cook2007concepts}.
These comparisons are shown in Fig.~\ref{figure2}\textbf{b} using continuous lines and open circles for the results obtained using the FE and PWE formulations, respectively.
For the FE-based results, we also show a polarization metric given by $p(u_x, u_y, u_z) = \int_V u_z^2 \, dV / \int_V (u_x^2 + u_y^2 + u_z^2) \, dV$, where $V$ is the unit cell volume.

Considering the dispersion relations for the baseline case (without undulation, i.e., $d_0=0$), we notice (left panel in Fig.~\ref{figure2}\textbf{b}) the occurrence of distinct branches corresponding to the flexural (F), in-plane shear (S), torsional (T), and longitudinal (L), behavior.
The wave modes corresponding to each branch are indicated at the edge of the first Brillouin zone ($ka/\pi = 1$), using a red circle, blue diamond, red diamond, and blue circle, respectively, for the F, S, T, and L branches.
Since the dynamic equations of the coupled normal and transverse behavior are two-dimensional ($xz$-plane), the PWE method is only able to capture the flexural and longitudinal motions (i.e., branches whose wave modes are marked with circles). 

Next, we present the results obtained for $d_0/a = 0.01$ in Fig.~\ref{figure2}\textbf{b} (middle panel). 
In this case, it is possible to notice a veering effect~\cite{mace2012wave} between the longitudinal and flexural (L--F) branches, which transition from the longitudinal to the flexural motion, for the lower branch (and vice-versa, for the upper branch).
Indicative wave modes are also shown for $ka/\pi = 0.32$ and $ka/\pi = 0.60$ for the longitudinal (blue circle) and flexural wave modes (red circle), respectively.
A similar veering effect is also noticed between the torsional and shear branches, which however are not represented by the PWE formulation.
In this case, veering occurs below $\overline{\omega} = \pi/2$ ($a/\lambda_L < 1/4$), characterizing a sub-Bragg operation. We also show the dispersion relations obtained for $d_0/a = 0.10$, which present excellent agreement with the FE-based solution for $\overline{\omega} < \pi/2$ and relatively good agreement for $\overline{\omega} > \pi/2$. We also indicate a highly polarized longitudinal-flexural wave mode at $ka/\pi = 0.60$ (green circle).
The results yielded by the PWE and FE methods present excellent agreement, with small divergences due to the limitations of the analytical equations used for the PWE formulation, which consider smooth curvatures (i.e., $d_0 \ll a$).

To obtain the undulation height ($d_0/a$) necessary to convert longitudinal to flexural waves at a given frequency, we intersect the second propagative branches shown in Fig.~\ref{figure2}\textbf{a} with the isofrequency line corresponding to the desired operating frequency.
Consider the illustrative case of $\overline{\omega}=1$, shown in Fig.~\ref{figure2}\textbf{c} (black line) combined with the second propagative branch in the first Brillouin zone ($ka/\pi \in [-1,1]$).
At this frequency, the longitudinal and flexural modes propagating in the flat structure ($d_0/a=0$) are represented, respectively, using the blue and red circles (for both positive and negative wavenumbers).
The point of maximum of the isofrequency curve (as projected onto the $k \times \omega$ plane) is located at $d_0/a=0.093$, whose second propagative branch is highlighted as a green dashed curve and with a maximum indicated using a green circle.
It is also interesting to recall that distinct points present varying values of group velocity along the isofrequency line, which are computed and shown in Fig.~\ref{figure2}\textbf{d}, normalized with respect to the longitudinal wave velocity.
The group velocity starts at $c_g/c_L=1$, corresponding to purely longitudinal waves (blue circle), crosses the zero group velocity line $c_g/c_L=0$ at $d_0/a=0.093$ (green circle), and returns to $c_g/c_L = -0.41$ at $d_0/a=0$ (red circle), corresponding to negative-propagating flexural waves.
We note that the point of zero group velocity indicates the undulation height necessary for mode conversion.
Additionally, by employing a graded undulated profile, it is possible to achieve a smooth transition between longitudinal waves ($c_g>0)$ and the zero group velocity point ($c_g=0$), which then leads to the reflection as flexural waves ($c_g<0$).
However, it is important to note that such perfect conversion would require a graded structure with an infinite length to ensure a continuous transition between the group velocities of each contiguous unit cell, which is not realizable in practice.

\subsection{Experimental verification of band gaps}

\begin{figure*}
 \centering
 \includegraphics{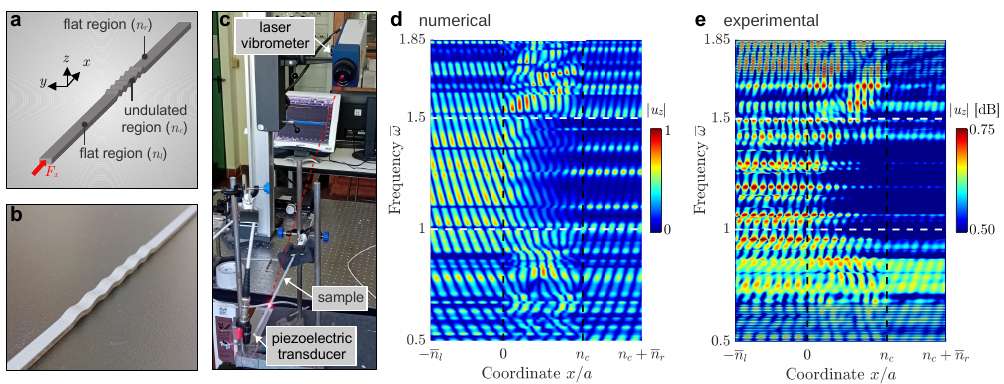}
 \caption{Time harmonic simulations.
 \textbf{a}. Simulated system with two flat regions (lengths $n_l \times a$ and $n_r \times a$) divided by an undulated region (length $n_c \times a$). A time harmonic longitudinal force ($F_x$) is applied at the left end of the beam.
 \textbf{b}. Manufactured test specimen shown close to the undulated region.
 \textbf{c}. Experimental setup showing the fabricated sample with an attached piezoelectric transducer resting on a soft support and the laser Doppler vibrometer used to perform the measurements.
 \textbf{d}. Simulation results for normalized transverse displacements along the beam axis and the frequency range $\overline{\omega} \in [0.5,1.85]$.
 The vertical dashed black lines indicate the undulated region ($x/a \in [0,n_c]$) and the horizontal dashed white lines indicate the band gap region. The colorbar represents absolute displacements.
 \textbf{e}. Experimental results corresponding to the numerical simulation in (\textbf{d}), with the colorbar adjusted for visualization.}
 \label{figure3}
\end{figure*}

To validate experimentally the computed dispersion relations, we verify the existence of a band gap in a beam having an undulated region with a profile described by $d_0/a=0.093$, corresponding to a band gap in the frequency range $\overline{\omega} \in [1.0,1.5]$.
We note that the ratio between the unit cell length and the longitudinal wavelength in this case is given by $a/\lambda_L = 1/2\pi \approx 0.16$. Thus, considering a structure composed of $6$ unit cells still allows to maintain subwavelength conditions ($6 \times a/\lambda_L \approx 0.95 < 1$). 
We then consider the structure represented in Fig.~\ref{figure3}\textbf{a}, composed of an undulated region with $n_c$ unit cells dividing two flat regions (i.e., $d_0/a=0$) with lengths $n_l \times a$ and $n_r \times a$.
Test specimens were then fabricated using Aluminum 5000 alloys (Al-Mg, EN AW 5056A) with a length of $300$ mm, due to fabrication limitations. The flat and undualted regions are defined by the values $n_l=14$, $n_c=6$, and $n_r=10$ unit cells.
An example of a fabricated specimen is shown in Fig.~\ref{figure3}\textbf{b}, zoomed at the undulated region.
For the experimental validation, we attached a piezoelectric transducer to one side of the test specimen. The specimen was clamped at the other end using a mechanical vise. A layer of low-density foam was inserted between the clamp and the structure to reduce transmitted vibrations and attenuate oscillations at the boundaries. The support is then mounted on a linear stage to measure transverse displacements along a $160$ mm length using a data acquisition system by using a fixed laser Doppler vibrometer. The setup is shown in Fig.~\ref{figure3}\textbf{c}.

To verify the existence of band gaps, a harmonic signal is applied using a longitudinal force at the left edge (equal for all FE nodes), performing a sweep in the frequency range $\overline{\omega} \in [0.5,1.85]$ (corresponding to $f \in [40,150]$ kHz) and computing displacements in the transverse direction at the top surface of the beam using free boundary conditions at the other faces.
Results from FE simulations are indicated in Fig.~\ref{figure3}\textbf{d}, with normalized displacements shown in absolute values in the range $[-\overline{n}_l, n_c+\overline{n}_r]$, with $\overline{n}_l = 5.5a$ and $\overline{n}_r = 4.5a$, chosen in accordance with experimental restrictions.
Within the band gap (marked with horizontal dashed white lines), standing waves are formed within the flat regions (to the left, $x/a < 0$, and to the right, $x/a > n_c$, of the undulated region), while a significant attenuation is observed within the undulated region ($x/a \in [0,n_c]$, marked using vertical dashed black lines) from left to right.

Experimental results are shown in Fig.~\ref{figure3}\textbf{e}, using a logarithmic scale for the resulting fast Fourier transforms (FFT) of the measured transverse displacements, restricting their values to the $[0.50,0.75]$ range to facilitate visualization. In this case, although the band gap is slightly shifted downwards in frequency, at around $\overline{\omega} \in [0.9,1.4]$ (i.e., an approximate $10\%$ decrease),
probably due to some uncertainty in the material properties,
this still represents a very good correlation with numerical data.
We also notice a particularly good correlation between numerical and experimental data in the corrugated region ($x \in [0,n_c]$) outside the band gap region, i.e., for $\overline{\omega}<1.0$ and $\overline{\omega}>1.5$.

Finally, we note that in the usual conversion mechanisms presented in literature, both types of waves present the same propagation direction.
In our proposed structures, however, longitudinal and flexural waves present an inversion of the propagation direction (i.e., opposing group velocity signs), enabling wave conversion through reflection.
Time-harmonic simulations do not allow for a clear appreciation of the wave conversion mechanism due to reflections of waves at the boundaries.
For this reason, in the next section we verify the wave conversion mechanisms using time-transient simulations and experiments.

\section{Wave conversion application} \label{sec_wave_conversion}

\subsection{Numerical investigation using a constant undulation profile}

\begin{figure*}[h!]
 \centering
 \includegraphics{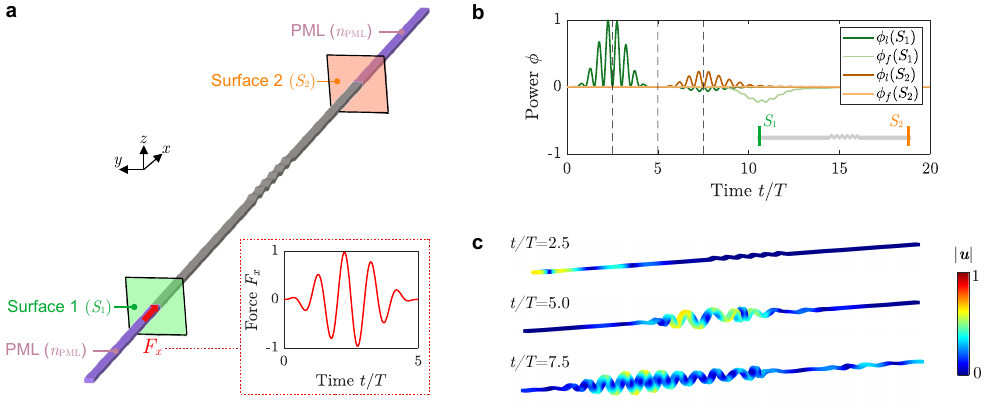}
 \caption{Numerical simulation of a longitudinal transient pulse applied to a beam with an undulated region.
 \textbf{a}. A longitudinal force ($F_x$) is applied using a modulated sinusoidal pulse (inset). 
 Surfaces $S_1$ and $S_2$ (in green and orange, respectively), located at the edges of the flat regions, are considered for the analysis of mechanical energy flux.
 Two PMLs (lengths $n_{\text{PML}} \times a$)  are included at the outer edges of each flat region.
 \textbf{b}. Longitudinal ($\varphi_l$) and flexural elastic power ($\varphi_f$) computed at the surfaces $S_1$ (green) and $S_2$ (orange). The sign indicates forward- ($>0$) or backward-propagating energy ($<0$). The dashed vertical lines indicate the time instants $t/T = 2.5$, $5.0$, and $7.5$, whose absolute normalized displacements are represented in (\textbf{c}).
 }
 \label{figure4}
\end{figure*}

We now proceed to verify numerically the longitudinal-flexural wave conversion functionality of the structure presented in the last section.
For the numerical time-transient simulations, two perfectly matched layers (PMLs), with a length of $n_{\text{PML}} \times a$ each, are added to both ends of the structure (purple region in Fig.~\ref{figure4}\textbf{a}) to minimize reflections.
We consider the FE implementation as described in~\cite{fathi2015time} with $n_{\text{PML}} = 15$.
Additionally, we consider two surfaces of interest, labeled as $S_1$ and $S_2$ (green and orange surfaces, respectively, represented in Fig.~\ref{figure4}\textbf{a}), which pass through the geometrical centers of the elements adjacent to each PML region, thus presenting a negligible distance to the PMLs ($\approx 0.25$ mm) with respect to the wavelength.
These surfaces are used to investigate the displacements and mechanical energy flux, 
in which case the evaluation of internal stresses at Gauss points are required to minimize numerical errors associated with the FE implementation~\cite{cook2007concepts}.
We then apply a longitudinal force following a signal with central frequency $\overline{\omega}=1$ (fundamental period $T = 1/f = 12.34$ $\mu$s) and normalized amplitude considering $5$ pulses modulated by a Hanning window (inset of Fig.~\ref{figure4}\textbf{a}). 
The time-transient simulations are performed considering a time-step incremental analysis using the Extended Newmark Method~\cite{fathi2015time}, with a total time duration of $t/T=20$ and an incremental time step $\Delta_t = 0.5$ $\mu$s ($T/\Delta_t \approx 25$).

In order to quantify the efficiency of wave conversion, we analyze it from a quantitative point of view by computing the corresponding elastic energy flux across surfaces $S_1$ and $S_2$.
The ratio of reflection or transmission between longitudinal and flexural waves are obtained using
\begin{equation} \label{eq_mechanical_energy}
\begin{aligned}
    R_f &= -\frac{\int \phi_f^-(S_1) \, dt }{\int \phi_l^+(S_1) \, dt } \, , \\
    R_l &= -\frac{\int \phi_l^-(S_1) \, dt }{\int \phi_l^+(S_1) \, dt } \, ,  \\
    T_l &= \frac{\int \phi_l^+(S_2) \, dt }{\int \phi_l^+(S_1) \, dt } \, , 
\end{aligned}
\end{equation}
where $\phi_l(S_i)$ ($\phi_f(S_i)$) is the instantaneous longitudinal (flexural) power (surface integral of the elastic energy flux) at the surface $S_i$, $i = \{1,2\}$, and the $+$ ($-$) superscript denotes energy flux in the positive (negative) direction.
The derivation of these fluxes is given in Appendix~\ref{app_mechanical}.

These quantities are computed for surfaces $S_1$ and $S_2$ and shown (normalized) in Fig.~\ref{figure4}\textbf{b}.
We notice that $\phi_l(S_1)$ presents an inversion in sign due to its reflection at the interface between the flat and undulated regions.
The transmitted power of longitudinal waves $S_2$ and reflected power of flexural waves $S_1$, indicated, respectively, as $\phi_l(S_2)$ and $\phi_f(S_1)$ are also shown. 
The reflection (flexural and longitudinal waves) and transmission (longitudinal waves) coefficients are respectively computed as $R_f = 0.56$, $R_l = 0.09$, and $T_l = 0.33$, thus demonstrating that most of reflected energy is converted in the form of flexural waves. Also, a non-negligible small energy fraction is reflected in the form of longitudinal waves due to the impedance mismatch between the undulated and the flat regions.
We also note that, although the value of $R_f$ does not seem particularly high, the ratio of peak absolute values between transverse and longitudinal displacements at $S_1$ is equal to $0.92$ for the beam centerline (see~ Appendix\ref{app_displacements}), which indicates that incident longitudinal waves can be used to generate flexural waves with a similar amplitude.

The corresponding computed displacements are illustrated in Fig.~\ref{figure4}\textbf{c} for increasing time instants $t/T= 2.5$, $5.0$, and $7.5$, with absolute normalized values represented in the colour scale.
It becomes clear that an incident longitudinal wave is mostly reflected as flexural waves and marginally transmitted as longitudinal waves. 

For the sake of comparison, we perform the same analyses for the operating frequency $\overline{\omega} = 0.85$.
The ratio $d_0/a=0.18$ is obtained as the point of maximum for the curve intersecting the surface corresponding to the second branch at this frequency.
In this case, the ratio between the unit cell length and the longitudinal wavelength is given by $a/\lambda_L = 0.85/2\pi \approx 0.14$, which can accommodate $n_c = 7$ unit cells and still remain in the subwavelength operating range ($7 \times a/\lambda_L \approx 0.95 < 1$).
We also set $n_l=13$, keeping $n_r = 10$.
We omit the graphical representation of these results for the sake of brevity.
Instead, we only present here the transmission and reflection coefficients, which are obtained as $R_f = 0.59$, $R_l = 0.26$, and $T_l = 0.10$.
Thus, there is an improvement in the flexural reflection coefficient due to the increase in both the $d_0/a$ ratio and the number of unit cells.
There is also an increase in the reflection coefficient and a decrease in the transmission coefficient of flexural waves, both owing to the larger impedance mismatch between the flat and undulated regions. 

It is important to note that, although rigorous, the previous analyses perform numerical integrations of stresses and velocities on the beam cross-section, which is not experimentally feasible. From a practical point of view, we are mostly interested in measurable quantities, such as displacements and velocities.
For this reason, we proceed to validate our findings experimentally by analyzing the time history of longitudinal and transverse displacements on the top surface of the beam.

\subsection{Experimental validation using a constant undulation profile}

\begin{figure*}[h!]
 \centering
 \includegraphics{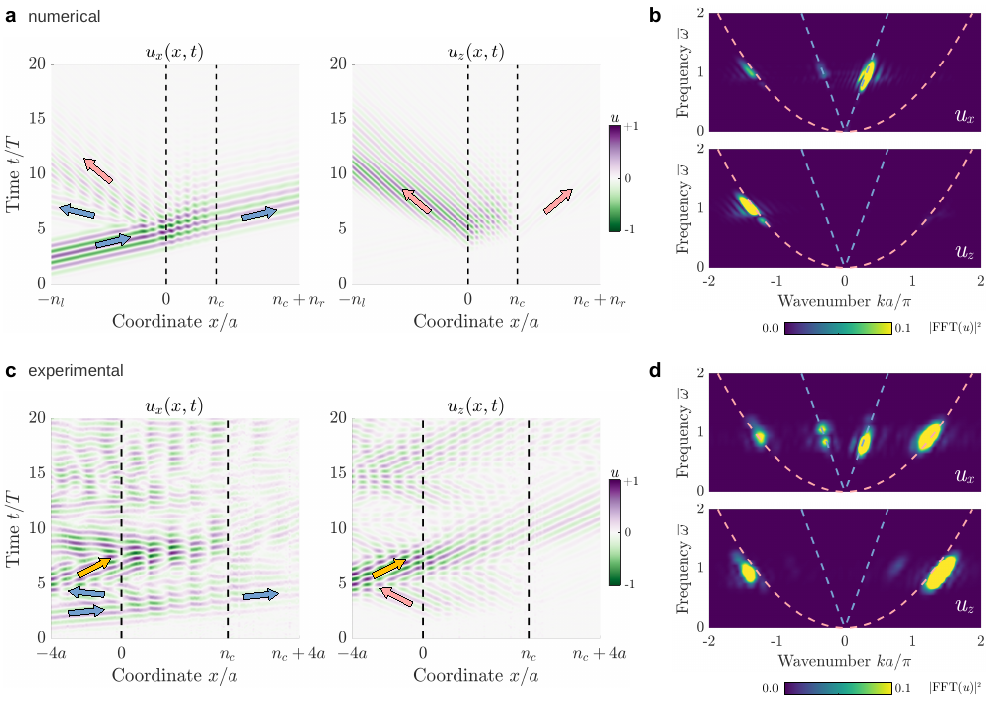}
 \caption{Numerical simulation and experimental validation of longitudinal to flexural wave conversion using a beam with a constant undulation profile.
 \textbf{a}. Computed longitudinal (left panel, $u_x$) and transverse (right panel, $u_z$) displacements at the top surface of the beam.
 Vertical dashed lines indicate the undulated region.
 Blue and red arrows represent, respectively, longitudinal and flexural wavefronts.
 The colorbar indicates normalized displacements
 \textbf{b}.2D-FFTs corresponding to the longitudinal ($u_x$, top panel) and transverse ($u_z$, bottom panel) displacements shown in (\textbf{a}). 
 Blue and red dashed lines indicate, respectively, longitudinal and flexural analytical dispersion branches at the flat region.
 The colorbar indicates normalized 2D-FFT values.
 (\textbf{c},\textbf{d}). Same as (\textbf{a},\textbf{b}), obtained experimentally.
 }
 \label{figure5}
\end{figure*}

Fig.~\ref{figure5}\textbf{a} shows the computed normalized longitudinal and transverse displacements using normalized spatial ($x/a$) and time coordinates ($t/T$),
with dashed black lines indicating the undulated region ($x/a \in [0,n_c]$).
The longitudinal displacements ($u_x$, left panel) indicate an impinging longitudinal wave (bottom left blue arrow) being partially reflected at the interface between the flat and the undulated regions (top left blue arrow). In addition, part of the energy is converted as reflected flexural waves (red arrow), while another part is transmitted as longitudinal waves (right blue arrow). 
Although these refer to longitudinal displacements, the distinction between longitudinal and flexural waves can be made due to the differences in propagation velocities. 
Likewise, the transverse displacements ($u_z$, right panel) indicate a wavefront associated with reflected flexural waves (red arrow on the left), while the transmitted flexural waves (red arrow on the right) are barely noticeable.
The transverse displacements associated with the Poisson's effect (i.e., due to longitudinal displacements) are negligible and cannot be distinguished. 

We also perform two-dimensional fast Fourier transforms (2D-FFTs) on these data, obtaining a representation in the wavenumber-frequency space ($k \times \omega$).
The numerical data are zero-padded (total of $2^{11}$ points) to improve the FFT resolution.
The obtained surfaces are represented in Fig.~\ref{figure5}\textbf{b} with the normalized coordinates $ka/\pi$ and $\overline{\omega}$, using normalized absolute FFT values.
We also plot, for reference, the analytical dispersion curves for longitudinal and flexural waves, respectively, using dashed blue and red curves.
The observations concerning the time histories are confirmed by the 2D-FFT, where it is possible to notice the existence of incident ($k>0$) longitudinal waves at $\overline{\omega}=1$ (top panel, blue curve), which are reflected ($k<0$) as longitudinal (blue curve) and flexural waves (red curve).
Likewise, the 2D-FFT of transverse displacements (bottom panel) indicate that the reflected flexural waves ($k<0$, red curve) are dominant with respect to transmitted fexural waves ($k>0$).

An analogous analysis is then performed considering data obtained experimentally using the setup presented in Fig.~\ref{figure3}\textbf{c}.
To this end, the vibrometer performs two sets of measurements over a $200$ mm length, using (i) a grazing angle to obtain longitudinal displacements ($u_x$) and (ii) a normal angle to obtain transverse displacements ($u_z$).
Fig.~\ref{figure5}\textbf{c} presents these results ($u_x$, left panel, and $u_z$, right panel) for the time interval $t/T \in [0,20]$.
We note that due to positioning errors associated with the measured area (which is smaller than the beam length), we only represent a region corresponding to the obtained data which is symmetric with respect to the undulations (i.e., centered at $x/a = n_c/2$).
The measured longitudinal displacements indicate the expected incident (bottom left blue arrow), transmitted (right blue arrow), and reflected longitudinal wavefronts (top left blue arrow).
We also observe an additional flexural incident wavefront (yellow arrow), which can be distinguished due to its group velocity, which is smaller than the one associated with longitudinal waves, as revealed by the inclination of the arrow.
The existence of this incident flexural wave is caused by the spurious excitation of a flexural mode caused by the piezoelectric transducer attached longitudinally to the test specimen.
Nonetheless, the wave conversion phenomenon can still be observed when analyzing the transverse displacements, where it is possible to notice reflected (left red arrow) and transmitted flexural waves (right red arrow).
In this case too, the incident flexural wave is observed (yellow arrow), which possesses the same group velocity as the wave indicated by the red arrow with an opposing propagation direction.
For $t/T > 10$, additional negative-going flexural waves are observed due to the subsequent incident longitudinal waves that impinge upon the undulated region after reflecting at the left edge of the test specimen.

The representation of this temporal data in the reciprocal space is obtained by performing a 2D-FFT on the displacements time history with an additional Hanning window in the time domain $t/T \in [0,10]$ to minimize the effect of additional reflections. The resulting surfaces are also zero-padded to improve the FFT resolution.
These results are shown in Fig.~\ref{figure5}\textbf{d}.
Concerning longitudinal displacements ($u_x$, top panel), the main differences with respect to the numerical simulation (see Fig.~\ref{figure5}\textbf{b}) are increases
(i) in the occurrence of backward-propagating waves ($k<0$) caused by reflections at the right edge of the specimen and
(ii) in the components associated with forward-propagating flexural waves (red dashed curve, $k>0$) caused by the spurious mode excitation. In both cases, these effects are not observed in the numerical simulation due to the perfect longitudinal excitation and the presence of PMLs to reduce internal reflections.
On the other hand, the main difference between numerical and experimental results ($u_z$, bottom panel) is the occurrence of forward-propagating flexural waves (red dashed curve, $k>0$) caused by the spurious mode excitation. Also, due to the reduced time window considered for the 2D-FFT, the backward-propagating flexural waves (red dashed curve, $k<0$) are associated with the conversion between longitudinal and flexural waves, thus demonstrating the wave conversion mechanism.

We conclude by noting that the use of a constant undulation profile yields significant reflection of longitudinal waves. Therefore, we next evaluate the use of a graded undulation profile as an alternative to reduce the reflection of longitudinal waves.

\subsection{Numerical investigation using a graded profile}

\begin{figure*}[h!]
 \centering
 \includegraphics{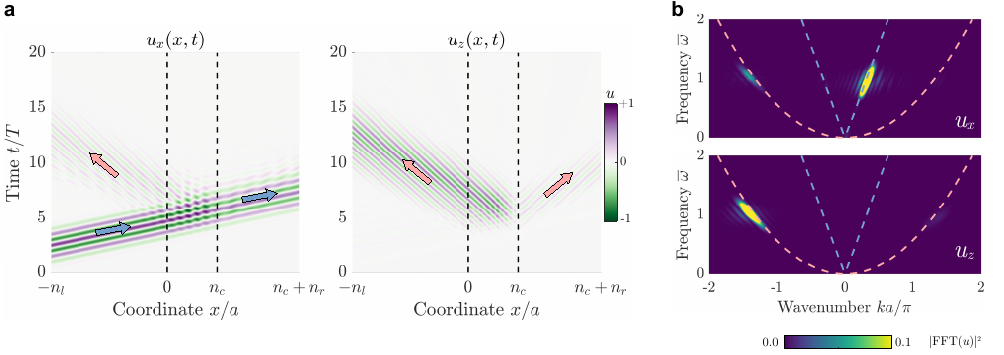}
 \caption{Numerical simulation of longitudinal to flexural wave conversion using a beam with a graded undulation profile.
 \textbf{a}. Computed longitudinal (left panel, $u_x$) and transverse (right panel, $u_z$) displacements at the top surface of the beam.
 Vertical dashed lines indicate the undulated region.
 Blue and red arrows represent, respectively, longitudinal and flexural wavefronts.
 The colorbar indicates normalized displacements.
 \textbf{b}. 2D-FFTs corresponding to the longitudinal ($u_x$, top panel) and transverse ($u_z$, bottom panel) displacements shown in (\textbf{a}). 
 Blue and red dashed lines indicate, respectively, longitudinal and flexural analytical dispersion branches at the flat region.
 The colorbar indicates normalized 2D-FFT values.}
 \label{figure6}
\end{figure*}

The same numerical analysis is now performed considering a graded undulated profile,  
where $d_0/a$ presents a linear variation up to $0.093$, distributed along $6$ unit cells in the undulated region.
The displacement history for longitudinal and transverse directions is presented in Fig.~\ref{figure6}\textbf{a}.
Concerning the longitudinal displacements ($u_x$, left panel), it is interesting to notice that no reflected longitudinal waves (blue arrows) are distinguishable.
On the other hand, the time history of transverse displacements ($u_z$, right panel) indicates that the peak displacement values are achieved for larger time instants with respect to the beam with a constant undulated profile (when compared to the constant profile, see Fig.~\ref{figure5}\textbf{a}, right panel), which is explained by the gradual modulation of group velocity inside the undulated region.
These results are confirmed by their 2D-FFTs, shown in Fig.~\ref{figure6}\textbf{b}, which show that the reflection of longitudinal waves is barely visible ($u_x$, top panel).
The ratio between the peak values for the transverse and longitudinal displacements is computed as $0.37$, indicating a decrease with respect to the constant undulation profile.

Finally, we compute the reflection and transmission coefficients, omitting the curves corresponding to mechanical power for the sake of brevity.
The reflection (flexural and longitudinal waves) and transmission (longitudinal waves) coefficients are computed as $R_f = 0.41$, $R_l \approx 0$, and $T_l = 0.55$.
We notice that the reflection coefficient for longitudinal waves is in good agreement with the 2D-FFT results.
These results also demonstrate that the graded profile presents a decrease in performance with respect to flexural wave conversion when a short undulation length is considered, which is compensated, however, by the decrease in reflection of longitudinal waves.
This design option thus provides a viable means of producing significant levels of flexural waves through the incidence of longitudinal waves with negligible reflection of longitudinal waves.

\section{Conclusions} \label{sec_conclusions}

In this work, we have proposed the use of phononic crystals for longitudinal-flexural wave conversion applications using beams with periodic and graded undulated profiles. 
The design of these systems is performed by analyzing the dispersion relation using a semi-analytical (PWE) method based on the dynamic equations relating longitudinal and transverse motion, also considering the effects of shear strain and rotational inertia.
These dispersion relations present an excellent agreement with equivalent FE-based solutions.
The results demonstrate that the curvature of the beam profile (i.e., undulation) can be optimized to obtain veering between the otherwise decoupled longitudinal and flexural branches of the dispersion diagram.
Such veering promotes a shift from longitudinal to flexural behavior within the same branch, which occurs in a sub-Bragg regime.

This phenomenon can be exploited to design waveguides whose length is similar to the wavelength of longitudinal waves in the flat region of the material. For a given operating frequency, a finite sample is proposed and its computed band gap experimentally demonstrated using a test specimen fabricated in Aluminum.
We then investigate the wave conversion capabilities of this waveguide with an undulated region of constant undulation profile using a longitudinal force applied as a modulated pulse and showing that most of the energy is reflected as flexural waves. Also, a small portion of the longitudinal wave energy is reflected due to the impedance mismatch between the flat and undulated regions. This behavior is numerically and experimentally demonstrated, confirming the wave conversion mechanism in the wavenumber-frequency space, also demonstrating that a graded profile is not strictly necessary to achieve wave mode conversion.
Finally, a beam with a graded undulated region is proposed to minimize the reflection of longitudinal waves, which is practically reduced to zero.
As a consequence of the smaller number of unit cells with the target undulation value, the conversion to flexural waves is also diminished, although remaining significant.

Interesting directions for future investigations are the functional behavior of corrugated surfaces in biological systems such as shells and their bio-inspired applications~\cite{liu2024seashell}.
Likewise, the application of corrugated surfaces to achieve wave conversion for large scale vibration protection applications and the creation of metasurfaces for topological waveguiding are additional promising directions.

\bibliographystyle{elsarticle-num} 
\bibliography{biblio}

\clearpage
\appendix
\setcounter{figure}{0}

\section{Eigenproblem matrices} \label{app_eigenproblem_matrices}

The terms of matrix $\mathbf{D}$ in Eq.~(\ref{eq_pwe1}) are given by
\begin{equation} \label{eq_pwe2}
\footnotesize
\begin{aligned}
 \mathbf{D}_{xx}^{(ij)} &= \frac{E}{\rho} (k+G_j)^2 \delta_{i,j} +
 \frac{\alpha \mu}{\rho}
 d_0^2 \bigg(\frac{\pi}{a}\bigg)^4 \bigg( \delta_{i-j,2} + 2 \delta_{i,j} + \delta_{i-j,-2} \bigg) \, , \\ 
 \mathbf{D}_{xz}^{(ij)} &= - \frac{ E + \alpha \mu }{\rho}
 d_0 \bigg(\frac{\pi}{a}\bigg)^2 \bigg( \delta_{i-j,1} + \delta_{i-j,-1} \bigg) 
 \text{i} (k+G_j) \nonumber \\
 &- \frac{E}{\rho} 
 d_0 \bigg(\frac{\pi}{a}\bigg)^3 \bigg( \delta_{i-j,1} - \delta_{i-j,-1} \bigg) \, , \\
 \mathbf{D}_{xy}^{(ij)} &= \frac{\alpha \mu}{\rho} 
 d_0 \bigg(\frac{\pi}{a}\bigg)^2 \bigg( \delta_{i-j,1} + \delta_{i-j,-1} \bigg) \, , \\ 
 \mathbf{D}_{zx}^{(ij)} &= \frac{ E + \alpha \mu }{\rho}
 d_0 \bigg(\frac{\pi}{a}\bigg)^2 \bigg( \delta_{i-j,1} + \delta_{i-j,-1} \bigg) 
 \text{i}(k+G_j) \nonumber \\
 &+ \frac{ \alpha \mu }{\rho} 
 d_0 \bigg(\frac{\pi}{a}\bigg)^3 \bigg( \delta_{i-j,1} - \delta_{i-j,-1} \bigg) \, , \\ 
 \mathbf{D}_{zz}^{(ij)} &= \frac{ \alpha \mu }{\rho} (k+G_j)^2 \delta_{i,j} +
 \frac{E}{\rho}
 d_0^2 \bigg(\frac{\pi}{a}\bigg)^4 \bigg( \delta_{i-j,2} + 2 \delta_{i,j} + \delta_{i-j,-2} \bigg)  \, , \\
 \mathbf{D}_{zy}^{(ij)} &= \frac{ \alpha \mu}{\rho} \text{i}(k+G_j) \delta_{i,j} \, , \\ 
 \mathbf{D}_{yx}^{(ij)} &= \frac{A}{I} \frac{\alpha \mu}{\rho} 
 d_0 \bigg(\frac{\pi}{a}\bigg)^2 \bigg( \delta_{i-j,1} + \delta_{i-j,-1} \bigg) \, , \\ 
 \mathbf{D}_{yz}^{(ij)} &= - \frac{A}{I} \frac{\alpha \mu}{\rho} \text{i}(k+G_j) \delta_{i,j} \, , \\
 \mathbf{D}_{yy}^{(ij)} &= \frac{E}{\rho} (k+G_j)^2 \delta_{i,j} + \frac{A}{I} \frac{ \alpha \mu }{\rho} \delta_{i,j} \, .
\end{aligned}
\end{equation}

The definition of material and geometrical properties is described in the main text.

\section{Mechanical energy flux and power} \label{app_mechanical}

The instantaneous mechanical energy flux at a given point is computed as
\begin{equation}
    I_i = -\sum_j \sigma_{ij} v_j \, ,
\end{equation}
where $v_j$ is the particle velocity in the $j$-th direction ($j=\{x,y,z\}$), with stresses and velocities presenting spatial ($x,y,z$) and temporal ($t$) dependence, omitted for the sake of simplicity.
For our proposed system, $\tau_{xy} \approx 0$ , leading to $I_x = -\sigma_x v_x - \tau_{xz} v_z$. By rewriting normal stresses and velocities in terms of their mean ($\overline{\bullet}$) and variable values ($\tilde{\bullet}$) with respect the cross-section neutral line ($z=0$), respectively, as $\sigma_x = \overline{\sigma}_x + \widetilde{\sigma}_x$ and 
$v_x = \overline{v}_x + \widetilde{v}_x$, the $x$-direction mechanical energy flux can be rewritten as
\begin{equation} \label{eq_long_power1}
I_x = 
- \overline{\sigma}_x \overline{v}_x
-( \overline{\sigma}_x \widetilde{v}_x + 
\widetilde{\sigma}_x \overline{v}_x ) 
- \widetilde{\sigma}_x  \widetilde{v}_x
 - \tau_{xz} v_z \, .
\end{equation}

Integrating this quantity over the surface $S$ of a cross section of the beam at any $x$-coordinate yields the instantaneous mechanical power
\begin{equation} \label{eq_mechanical_power}
    \int_S \, I_x(x,y,z,t) \, dS = 
    \phi_l(x,t) + \phi_f(x,t) \, ,
\end{equation}
where $\phi_l(x,t) = - \int_S \, \overline{\sigma}_x \overline{v}_x \, dS$ is the mechanical power associated with purely longitudinal waves and
$\phi_f(x,t) = - \int_S \, ( \widetilde{\sigma}_x  \widetilde{v}_x + \tau_{xz} v_z ) \, dS$  represents the analogous of flexural waves with a shear component. We note that $\int_S \, ( \overline{\sigma}_x \widetilde{v}_x + 
\widetilde{\sigma}_x \overline{v}_x ) \, dS = 0$, which represents a surface integral of  scalar fields with a linear distribution with respect to the neutral line.
For the sake of convenience, we adopt $\phi(x,t) = \phi(S)$, where $S$ is a surface at the $x$-coordinate and orthogonal to the $x$-axis.

\section{Displacements at the cross-section during wave conversion} \label{app_displacements}

\begin{figure}[h!]
 \centering
 \includegraphics{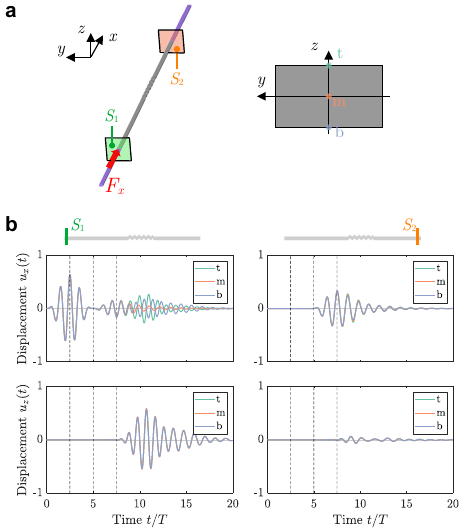}
 \caption{ \textbf{a}. displacements for the points at $y=0$ (top left inset): t (top, $z=+h/2$), m (middle, $z=0$), and b (bottom, $z=-h/2$) 
 \textbf{b}. Longitudinal ($u_x$, top panels) and transverse ($u_z$, bottom panels) displacements, shown at the surface $S_1$ (left panels) and $S_2$ (right panels). The dashed vertical lines indicate the time instants $t/T=2.5$, $5.0$, and $7.5$, }
 \label{figure_app1}
\end{figure}

We begin by analyzing the longitudinal ($u_x$) and transverse ($u_z$) displacements at the $S_1$ surface.
Three points of interest, located at $y=0$, are considered for illustrative purposes, namely, at the top (t, $z=+h/2$), middle (m, $z=0$), and bottom (b, $z=-h/2$) of the cross-section, as represented in Fig.~\ref{figure_app1}\textbf{a} (top left panel).
The displacements of these points are shown in the top and middle panels in Fig.~\ref{figure_app1}\textbf{b}, with dashed vertical lines indicating the time instants $t/T=2.5$, $5.0$, and $7.5$.
In a first time interval ($t/T \in [0,5]$), we notice that these points present the same longitudinal displacements, corresponding to a longitudinal wave (top left panel in Fig.~\ref{figure_app1}\textbf{b}).
In a second time interval ($t/T \in [5,8]$) we notice a similar behavior with a phase inversion, which corresponds to a reflection of longitudinal waves at the interface between the flat and undulated regions.
After a short instant of interaction between longitudinal and flexural waves ($t/T \in [8,9]$), we notice that the longitudinal displacements of point m are bounded by those of points b and t, corresponding, in fact, to their mean value, indicating a flexural behavior ($t/T \in [9,20]$).
This is also confirmed by the transverse displacements of the analyzed points (middle left panel in Fig.~\ref{figure_app1}\textbf{b}), which indicate a zero value of the m point for $t/T \in [0, 8]$, while points b and t present displacements with opposing signs produced by the Poisson's effect.
It is also possible to notice that the transverse displacements are very similar for $t/T \in [9,20]$, presenting small deviations with respect to those of the m point due to the propagation of flexural waves.
Considering surface $S_2$, the dominant behavior is shown to be associated with a longitudinal mode (top right panel in Fig.~\ref{figure_app1}\textbf{b}).
Additionally, the transverse displacements suggest a less significant flexural behavior, which indicates that a residual flexural energy is also transmitted for $t/T \in [7.5,20]$ (middle right panel in Fig.~\ref{figure_app1}\textbf{b}).

\end{document}